\documentstyle[emulateapj,psfig]{article}
\newcommand{\be}{\begin{equation}}
\newcommand{\ba}{\begin{eqnarray}}
\newcommand{\ee}{\end{equation}}
\newcommand{\ea}{\end{eqnarray}}  
\newcommand{\etal}{et al.\ }
\newcommand{\cmm}{\,{\rm cm}^{-2}}
\def\msun{\,{\rm M_\odot}}
\def\vir{{\rm vir}}
\def\Lya{Ly$\alpha$}
\def\CIV{\hbox{C$\scriptstyle\rm IV\ $}}
\def\SiIV{\hbox{Si$\scriptstyle\rm IV\ $}}
\def\OVI{\hbox{O$\scriptstyle\rm VI\ $}}
\def\nHI{{\rm HI}}
\def\nCIV{{\rm CIV}}
\def\nSiIV{{\rm SiIV}}

\def\HI{\hbox{H~$\scriptstyle\rm I\ $}}

\def\gtsima{$\; \buildrel > \over \sim \;$}
\def\ltsima{$\; \buildrel < \over \sim \;$}
\def\prosima{$\; \buildrel \propto \over \sim \;$}
\def\gsim{\lower.5ex\hbox{\gtsima}}
\def\lsim{\lower.5ex\hbox{\ltsima}}
\def\simgt{\lower.5ex\hbox{\gtsima}}
\def\simlt{\lower.5ex\hbox{\ltsima}}
\def\simpr{\lower.5ex\hbox{\prosima}}

\def\ie{{\frenchspacing i.e. }}
\def\eg{{\frenchspacing e.g. }}

\lefthead{Scannapieco, Ferrara, \& Madau}
\righthead{Early Enrichment of the IGM}
\submitted{submitted to the ApJ}
\makeatletter

\newenvironment{figurehere}
  {\def\@captype{figure}}
  {}
\makeatother

\begin{document}      

\title{Early Enrichment of the Intergalactic Medium and its Feedback on Galaxy
Formation}

\author{
Evan Scannapieco\altaffilmark{1}, Andrea Ferrara\altaffilmark{1}, \&
Piero Madau\altaffilmark{1,2}}

\altaffiltext{1}{Osservatorio Astrofisico di Arcetri, Largo E. Fermi 5, 50125 
Firenze, Italy.}
\altaffiltext{2}{Department of Astronomy and Astrophysics, University of
California, 1156 High Street, Santa Cruz, CA 95064.}

\begin{abstract}

Supernova--driven outflows from early galaxies may have had a large
impact on the kinetic and chemical properties of the intergalactic
medium (IGM).  We use three--dimensional Monte Carlo cosmological
realizations of a simple linear peaks model to track the time
evolution of such metal--enriched outflows and their feedback on
galaxy formation.  We find that at most 30\% of the IGM by volume is
enriched to values above $10^{-3} Z_\odot$ in models that only include
objects that cool by atomic transitions. The majority of enrichment
occurs relatively early ($5 \lsim z \lsim 12$) and leads to a
mass-averaged cosmological metallicity between $10^{-3} Z_\odot$ and
$10^{-1.5} Z_\odot$. The inclusion of Population III objects that
cool through ${\rm H}_2$ line emission has only a minor impact on
these results: increasing the mean metallicity and filling factor by
at most a factor of 1.4, and moving the dawn of the enrichment epoch
to $z \approx 14$ at the earliest. Thus enrichment by outflowing
galaxies is likely to have been incomplete and inhomogeneous, biased
to the areas near the starbursting galaxies themselves. Models with a
$10\%$ star formation efficiency can satisfactorily reproduce the
nearly constant ($2\leq z\leq 5$, $Z\approx 3.5\times 10^{-4}\,Z_\odot$) 
metallicity of the low column density Ly$\alpha$ forest derived by
Songaila (2001), an effect of the decreasing efficiency
of metal loss from larger galaxies. Finally, we show that IGM
enrichment is intimately tied to the ram-pressure stripping of baryons
from neighboring perturbations. This results in the suppression of at 
least 20\% of the dwarf galaxies in the mass range $\sim 3\times 
10^8-3\times 10^9\,M_\odot$ in all models with filling factors greater
than 2\%, and an overall suppression of $\sim 50\%$ of dwarf
galaxies in the most observationally-favored model.
\end{abstract}



\section{Introduction}

Recent QSO absorption line observations have shown that the
intergalactic medium (IGM) is polluted with heavy elements at
intermediate redshifts (Songaila \& Cowie 1996). From such
measurements of column density ratios $N_{\nCIV}/N_{\nHI}$, Hellsten
et al.\ (1997) and Rauch, Haehnelt, \& Steinmetz (1997) concluded that
typically [C/H]~$ \simeq -2.5$ at $z \simeq 3$, with an order of
magnitude dispersion about this mean value.\footnote{In the usual
notation, [C/H] = log (C/H) $-$ log (C/H)$_{\odot}$.}  These values,
however, refer to overdense regions of the universe, traced by \Lya\
clouds with column densities in excess of $\log N_\nHI = 14.5$.

The presence of metals has more recently been assessed in clouds in
which $\log N_\nHI < 14.0$ as reviewed by Pettini \etal (2001). At
these low optical depths, statistical techniques to extend the search
for highly ionized species such as \CIV and \OVI must be applied. The
results show that {\it i)} unrecognized weak \CIV systems must be
present in order to reproduce the full \CIV optical depth (Ellison
\etal 2000), and {\it ii)} that metals, as traced by \OVI are present
in a gas with a density lower than that of the mean IGM (Schaye \etal
2000). Very recently, Songaila (2001) has been able to trace the IGM
metallicity evolution in systems with $\log \CIV>12$ 
and conclude that a minimum metallicity $Z\approx 3.5\times
10^{-4} Z_\odot$ is already in place at $z=5$. 

Although these techniques help to extend QSO absorption studies to
underdense regions of the IGM, present observations are only able to
place a lower limit on the total volume filling factor of metals.
Current measurements, combined with numerical simulations, indicate
that metals associated with $\log N_\nHI\simlt 14.2$ filaments fill
$\simgt 3\%$ of intergalactic space, including areas far away from the
high overdensity peaks where galaxies form (Madau, Ferrara, \& Rees
2001, hereafter MFR).  This suggests that metal pollution occurred
relatively early, resulting in a more uniform distribution and
enriching vast regions of intergalactic space.  This allows the \Lya\
forest to be hydrodynamically `cold' at low redshifts, as
intergalactic baryons have time to relax again under the influence of
dark-matter gravity.  Note that the presence of high-redshift metals
is of great observational importantance, as the measurement of metal
lines in $z \gsim 6$ quasars may also serve as a probe of
reionzation (Oh 2002).

These observations prompted MFR to suggest high-redshift ($z \approx
10$) galaxy outflows as a mechanism for IGM enrichment.  This study
could not determine the metal filling factor produced in such a
scenario, however, as it was focused on the evolution of typical
objects at a single mass-scale and formation redshift.  Similar
outflow models have been proposed by Nath \& Chiba (1995) and
Scannapieco \& Broadhurst (2001, hereafter SB), but primarily
motivated by the chemical and thermal properties of the X-ray emitting
gas in galaxy clusters.  While the latter of these studies included a
range of galaxy masses and was able to make some estimates as to the
total filling factor of metals, these results were fairly crude as the
study was focused on the properties of individual galaxies.  Aguirre
et al.\ (2001a) and Aguirre et al.\ (2001b) studied IGM metal
enrichment by superimposing an outflow model on numerical simulations
that did not include SN-driven winds, but were only able to constrain the
contribution from late-forming ($z \lsim 6$) and relatively large ($M
\gsim 10^{8.5} M_\odot$) objects. Cen \& Ostriker (1999) studied
metal enrichment in even lower resolution smoothed particle
hydrodynamic (SPH) simulations with a dark matter particle mass of
$8.6 \times 10^8 M_\odot.$ Gnedin \& Ostriker (1997) studied the
relationship between reionization and early metal enrichment in
high-resolution simulations, but did not adequately follow supernova
explosions.  Finally, Thacker, Scannapieco, \& Davis (2002) were able
to estimate the filling factor of outflows at $z \geq 4$ purely in the
context of high-resolution SPH simulations with a dark matter particle
mass of $2.5 \times 10^6 M_\odot$, but were not able to examine its
dependence on model parameters due to the high computational cost of
this approach.

Early-enrichment scenarios also have important implications for the
thermal and velocity structure of the IGM, as first studied in
Tegmark, Silk, \& Evrard (1993) and Voit (1996) (see also Cen and
Bryan 2000). The resulting feedback on galaxy formation was first
examined in Scannapieco, Ferrara, \& Broadhurst (2000), SB, and
Scannapieco, Thacker, \& Davis (2001). The nature of this effect is
twofold: an impinging wind may shock-heat the gas of a nearby
perturbation to above the virial temperature, thereby mechanically
evaporating the gas, or the gas may be accelerated to above the escape
velocity and stripped from the perturbation entirely.  The latter
channel is considerably more effective, because shock-heated clouds
that are too large to be stripped are able to radiatively cool within
a sound crossing time, thus limiting evaporation.  Note that this type
of feedback is fundamentally different from the one commonly adopted
in galaxy formation models, in which hot gas is produced by supernovae
in the parent galaxy.

In this paper we return to the issues of enrichment and feedback,
adopting a more complete approach that combines the detailed modeling
of a typical object as in MFR, with the more general spatially
dependent modeling described in SB.  In this way we are able
place constraints on the overall metal filling factor produced as well
as investigate the link between cosmic metal enrichment and the
feedback from outflows on galaxy formation.

The structure of the paper is as follows.  In \S2 and \S3 we describe
our semi-analytical simulations of galaxy formation with feedback and
IGM enrichment.  In \S4 we summarize the results of these simulations
and the constraints they place of the fraction of the universe
impacted by outflows; conclusions are given in \S5.

\section{A Linear Peaks Model of Galaxy Formation}

In order to determine the distribution of outflows as a function of
cosmic time, we use the linear peaks model described in detail
in SB.  Note that it is important not only to have a measure of the
overall number density of such objects, but also of their spatial
distribution, as high-redshift galaxies are expected to be highly
clustered both from theory (e.g., Kaiser 1984) and observations
(Giavalisco et al.\ 1998).

Using a standard fit to the Cold Dark Matter (CDM) power-spectrum
(Bardeen et al.\ 1986), we construct a 256$^3$ linear density field
spanning a (4 $h^{-1}$ Mpc)$^3$ cubic comoving volume, where $h$ is
the Hubble constant in units of 100 km~s$^{-1}$~Mpc$^{-1}$.  Based
mainly on the latest measurements of Cosmic Microwave Background (CMB)
anisotropies (eg.\ Balbi \etal 2000; Netterfield \etal 2001; Pryke
\etal 2002) and the abundance of galaxy clusters (Viana \& Liddle
1996), we focus our attention on a cosmological model with parameters
$h=0.65$, $\Omega_M$ = 0.35, $\Omega_\Lambda$ = 0.65, $\Omega_b =
0.05$, $\sigma_8 = 0.87$, and $n=1$, where $\Omega_M$,
$\Omega_\Lambda$, and $\Omega_b$ are the total matter, vacuum, and
baryonic densities in units of the critical density, $\sigma_8$ is the
mass variance of linear fluctuations on the $8 h^{-1}{\rm Mpc}$ scale,
and $n$ is the tilt of the primordial power spectrum.

This linear density field is convolved with spherical `top--hat'
window functions corresponding to nine different total masses, spaced
in equal logarithmic intervals from $3.0 \times 10^7\,\msun$ to $4.3
\times 10^{11}\,\msun$, and spanning the interesting range from
objects that lie close to the lower limit set by photoionization and
molecular cooling (e.g., Barkana \& Loeb 1999; Ciardi, Ferrara, \&
Abel 2000) to the most massive galaxies that host outflows.  We assume
that the formation of Population III (Pop III) objects, defined as
halos with virial temperatures below $10^4$~K, is completely
suppressed by photodissociation of hydrogen molecules by UV radiation
produced by nearby objects, and we study the impact of this assumption
in further detail below.

Using the elliptical collapse model of Sheth, Mo, \& Tormen (1999), we
identify all `virialized' peaks in the overdensity field, arrange them
in order of decreasing collapse redshift, and exclude all unphysical
objects collapsing within more massive, already virialized halos.
After collapse, we account for the finite gas cooling time using a
simple inside--out collapse model (White \& Frenk 1991; Somerville
1997).  The cooling gas initially relaxes to an isothermal
distribution at the virial temperature $T_{\rm vir}$ in a Navarro,
Frenk, \& White (1997, hereafter NFW) dark matter halo with
concentration parameter $c =5$, and with a uniform metallicity $Z$
calculated as described in the next Section.  In this model, the gas
within a radius $r_{\rm cool}$ cools due to radiative losses that
account for metallicity as tabulated by Sutherland and Dopita (1993),
and the gas outside this radius stays at the virial temperature of the
halo, with $r_{\rm cool}$ moving outwards with time.  This model is
described in further detail in SB.

When the total mass contained within $r_{\rm cool}$ equals the
object's baryonic mass, a new galaxy is assumed to form with a gas
mass $M_b=({\Omega_b/\Omega_M})M.$ In the smaller
halos at high redshift having $\log T_{\rm vir}<5.7$ (i.e.  masses
$M<2\times 10^{10}\,[(1+z)/10]^{-3/2}\,\msun$), rapid cooling by
atomic hydrogen and helium occurs on timescales much shorter than the
gas free--fall time, and infalling gas collapses to the center at the
free--fall rate rather than coming to hydrostatic equilibrium
(MFR). The supply of cold gas for star formation is then only limited
by the infall rate.

We assume a Salpeter initial mass function (IMF) with upper and lower
mass cut--offs equal to $M_u=120\,\msun$ and $M_l=0.1\,\msun$,
respectively. In this case, one SN occurs for every $\nu^{-1}\approx
136\,\msun$ of stars formed, releasing an energy of $E_0 =
10^{51}$~erg.  This mechanical energy is injected by SNe after a few
times $10^7\,$ yr: at this stage SN--driven bubbles propagate into the
halo quenching further star formation, and the conversion of cold gas
into stars is limited by the increasing fractional volume occupied by
SN remnants.  To be conservative, we do not consider the possibility
that very massive ($\approx 300 M_\odot$), metal-free stars might
contribute both to metallicity and energy input. Their effects have
been discussed in detail by Schneider \etal (2002) and Schneider,
Guetta \& Ferrara (2002) to which we refer to the reader for an
extensive description.

Before SN feedback occurs some fraction
$f_\star$ of the gas will be able to cool, fragment, and form stars.
As our formalism does not include local feedback effects, this star
formation efficiency must be considered as a free parameter of the
model.  Finally, a fraction $f_w$ of this energy will be channeled at
a constant rate into a galaxy outflow over a timescale span of $t_{\rm
OB} = 33$~Myr, ejecting gas into the IGM.

As we will show below (Fig.\ 1), the most efficient IGM pollutors are
objects with masses of a few times $10^8 M_\odot$, for which the
fraction of gas that can cool in a free-fall time is essentially unity
(MFR). Hence, this gas is readily available to be transformed into stars
on short timescales.  This justifies the prompt star formation
(starburst mode) approximation we have adopted, which therefore should
be appropriate to the aims of this study.

\section{Modeling Galaxy Outflows}

The outflows are modeled as spherical shells using a method that is
based on the approach described in SB, but with several important
refinements taken from MFR, Ferrara, Pettini \& Shchekinov (2000), and
Mori, Ferrara \& Madau (2001; hereafter MFM).  An outflow is driven
out of the galaxy by internal pressure and decelerated by inertia and
the gravitational pull of the dark matter halo, both estimated in the
thin shell approximation (Ostriker \& McKee 1988; Tegmark, Silk, \&
Evrard 1993).

The expansion of the shell, whose radius is denoted by $R_s$, is
driven by the internal energy, $E_b$, of the hot bubble gas.  The
pressure of such a gas (with adiabatic index $\gamma=5/3$) is
therefore $P_b=E_b/2\pi R_s^3$. Momentum and energy conservation yield
the relevant evolutionary equations:
\ba
\dot{R_s} &=& \frac{3 (P_b-P)}{\rho R_s}
	 - \frac{3}{R_s}(\dot{R_s} - HR_s)^2
		    - \Omega_M  \frac{H^2 R_s}{2}
	 - g_s,\nonumber \\
\dot{E_b} &=&  L(t) - 4 \pi R_s^2 \dot{R_s} P_b - L_c,
\ea
where the dots represent time derivatives, the subscripts {\it s} and
{\it b} indicate shell and bubble quantities respectively, $g_s\equiv
GM(R_s)/R_s^2$, and $\rho$ is the density of the ambient medium, taken
to be the halo gas density within the virial radius and the mean IGM
background density outside the virial radius.  These
equations reduce to those given in MFR in the regime in which the
Hubble expansion is negligible and reduce to those given in SB if the
NFW profile is replaced by a point mass and external pressure is
neglected.

The cooling rate, $L_c$, is assumed here to be dominated by inverse
Compton cooling off CMB photons (Ikeuchi \& Ostriker 1986), as gas
radiative processes are much less efficient in the low density $10^5$ K 
$\le T \le 10^8$ K gas that drives the outflows.  This approximation is
especially appropriate as the combined cooling processes produce
variations of less than a few percent on the final size of the bubble
(see Fig. 6 of MFR).

The mechanical luminosity of SNe is given by 
$$
L(t) = (f_w E_0) {\nu M_\star \over t_{\rm OB}} \Theta(t_{\rm OB}-t) 
$$
\be
= 8.26 \times 10^{33} \Theta(t_{\rm OB}-t) \left({\Omega_b\over
\Omega_M}\right) f_w f_\star M {~~~\rm erg~s}^{-1}.
\label{meclum}
\ee This assumption of a constant luminosity over the burst is most
accurate for the larger galaxies in our simulations, in which the
stochastic variations of $L(t)$ become smaller due to the larger
number of SNe.  We constrain $f_w$ by combining the overall efficiency
of 30\% derived for the $2\times 10^8 M_\odot$ object simulated by MFM
with the mass scaling derived in Ferrara, Pettini \& Shchekinov
(2000), which was obtained by determining the fraction of starburst
sites that can produce a blowout in a galaxy of a given mass.  Thus,
we choose $f_w(M) = 0.3\delta_B(M)/\delta_B(M=2\times 10^8 M_\odot)$
where
\be \delta_B(M)=
\cases{ 
1.0 & $\tilde N_t \leq 1$ \cr 
1.0 - 0.165 \, {\rm ln} (\tilde N_t^{-1}) &
 $1 \leq \tilde N_t \leq 100$ \cr 
[1.0 - 0.165 \, {\rm ln} (100)] \, 100 \, \tilde N_t^{-1}  & 
$ 100 \leq \tilde N_t$ \cr}
\label{eq:deltaB}
\ee 
where  $\tilde N_t \equiv 1.7
\times 10^{-7} (\Omega_b/\Omega_M) M/M_\odot$ is a dimensionless 
parameter that scales according to the overall number of SNe produced 
in a starburst, divided by the efficiency $f_\star$.

Within the virial radius a fixed fraction $f_m = 0.5$ of the gas
is swept into the shell, a value taken from the numerical
simulations described in MFM. In those
experiments it is seen that after blow--out, half of the initial mass
contained in the virial radius re--collapsed to the center as a result
of the multiple shell--shell interactions leading to the formation
of cold sheets. 
In this case the halo gas is assumed to virialize to an isothermal 
distribution and settle down to a density profile 
\be 
\ln \rho(r)= \ln ( f_{\rm m} \,
\rho_0) \,-\, {\mu m_p\over 2kT_{\rm vir}}[v_e^2(0)- v_e^2(r)] ,
\ee 
(Makino, Sasaki, \& Suto 1998), where the central, preburst
gas density $\rho_0$
is determined by the condition that the total baryonic mass fraction
within the virial radius is equal to the cosmic average, yielding $\rho_0 =
11052 \rho_{\rm crit} \Omega_b.$

Outside the virial radius the shells expand into the Hubble flow,
sweeping up all of the baryons in their path.  Finally, when outflows
slow down to the point that they are no longer supersonic, our
approximations break down, and the shell is possibly fragmented by
random motions. At this point we let the bubble expand with the Hubble
flow.

To calculate the cooling time of forming galaxies (see \S2), we use a
simple estimate of the metallicity of collapsed halos.  The metals in
each outflow are assumed to be evenly distributed within its radius,
adopting an average yield from each supernova of 2 $M_\odot$ (\eg
Nagataki \& Sato 1997), half of which is deposited in the outflow and
half of which remains in the galaxy.  Each collapsing object is then
assigned a mass in metals $M_Z$, taken to be zero initially and
modified by each outflow passing within its collapse radius, $r_{\rm
coll}$.  We assume that the fraction of metals which fall into the
collapsing halo from a given outflow is equal to the volume fraction
of the outflowing bubble which falls within the collapsing sphere.  In
this case, for each overlapping bubble, $M_Z$ is updated to
\be
M_Z \longrightarrow M_Z +
\frac{V_{\rm overlap}}{\frac{4 \pi}{3} r_{\rm coll}^3}
M^{\rm blast}_Z 
\label{eq:met}
\ee 
where $V_{\rm overlap}$ is the volume of intersection between the
outflow and the collapsing sphere.  By dividing this mass by the total
baryonic mass of the galaxy we can compute the initial metallicity of
the object, and thus the delay between collapse and star formation due
to cooling.  Note however, that this cooling time is only significant
at the highest mass scales, $\simgt 10^{10} M_\odot.$

Our model is also able to account for the inhibition of surrounding
low-mass galaxy formation (\ie negative feedback).  Scannapieco,
Ferrara, \& Broadhurst (2000) showed that the most important such
mechanism is `baryonic stripping,' whereby high--redshift galaxy
outflows strip the gas out of nearby overdense regions that would have
otherwise become low--mass galaxies.  Whenever a shock moves through
the center an overdense region that has not yet virialized,
we apply a simple check to determine if such stripping has occurred.

In each case we estimate the comoving radius of the overdense region
as $r_{M} (1+\delta_{\rm NL})^{-1/3}$ where $\delta_{\rm NL}$ is the
nonlinear overdensity of the region, which estimated from the
spherical collapse model as
\be
1 + \delta_{\rm NL} = \frac{9}{2}
\frac{(\theta - \sin \theta)^2}{(1 - \cos \theta)^3},
\label{eq:sphere}
\ee
where the collapse parameter $\theta$ is given by $(\theta - \sin
\theta)^{2/3} \pi^{-2/3} = D(z_{\rm cross})/D(z_c)$ where $z_{\rm
cross}$ is the redshift at which the shock moves through the center of
the objects, $z_c$ is its collapse redshift, computed from
the model of Sheth \& Tormen (2001), and $D(z)$ is the linear growth 
factor as a function of redshift.
We then inhibit galaxy formation if the shock has
sufficient momentum to accelerate the gas to its escape velocity:
\be \omega \, M_s \, \dot R_s \geq M_p \, v_e,
\label{eq:strip}
\ee 
where $\omega$ is the solid angle of the shell subtended by the
perturbation, $M_s$ is the baryonic mass swept up by the shock,
including both the IGM and gas ejected from the host objects, $v_s$
the velocity of the shock, and $M_p$ and 
$v_e = 2 G M_p (1+\delta_{\rm NL})^{1/3}/r_{M_p}$ are the gas mass and
escape velocity of the perturbation.

In SB we simply estimated the filling factor of outflows by summing
over the volumes contained within all galaxy outflows, regardless of
their mass scales or spatial distribution.  As this investigation is
focused on determining the overall state of the IGM, here we adopt a
more accurate approach in which we construct a $256^3$ grid, search
through the expanding bubbles, and count all grid points within them
as a function of redshift.

Our approach also differs from that in SB in our treatment of outflows
expanding from host galaxies that merge.  While our previous estimates
only counted outflows from unmerged host galaxies, in this work we
instead take the wind to remain fixed at the comoving radius found at
the time of the merger.  This is because, while it is clear that our
approximation of a spherical pressure-driven wind is broken when a
merging galaxy passes through this shell, it is unlikely that such a
merger would be able to absorb all the gas ejected, somehow erasing
the wind from existence.

It is important to point out that although our estimate of
the filling factor takes into account the spatial orientation of the
bubbles, our formalism fails to capture the interaction between
expanding shock fronts.  In the case in which two outflows are
expanding in opposite directions, our approach allows them to pass
through each other unhindered, leading to a slight overestimate of the
volume impacted.  Similarly, in the case in which an outflow expands
into a bubble evacuated by a previous wind, it is nevertheless modeled
as if it were expanding into the Hubble flow, leading to a slight
underestimate.

\section{Results}

\subsection{Cosmic Metal Filling Factor}

In the left panels of Figure \ref{fig:vol}, we show the derived
filling factor of outflows.  Here, the bend at $z \approx 13$ is due
to our finite mass resolution. At this redshift, the smallest mass
scale drops below the virial temperature limit of $10^4$ K, causing
the effective resolution of the simulation to jump to the next mass
scale.  In reality objects between these two mass scales are formed,
and the metals from these objects would smooth the
evolution of the filling factor.  Finally, because our method becomes
inaccurate when perturbations are overly nonlinear, we restrict this
figure to redshifts above 3.

\begin{figurehere}
\centerline{
\hspace{+0.4cm}
\psfig{file=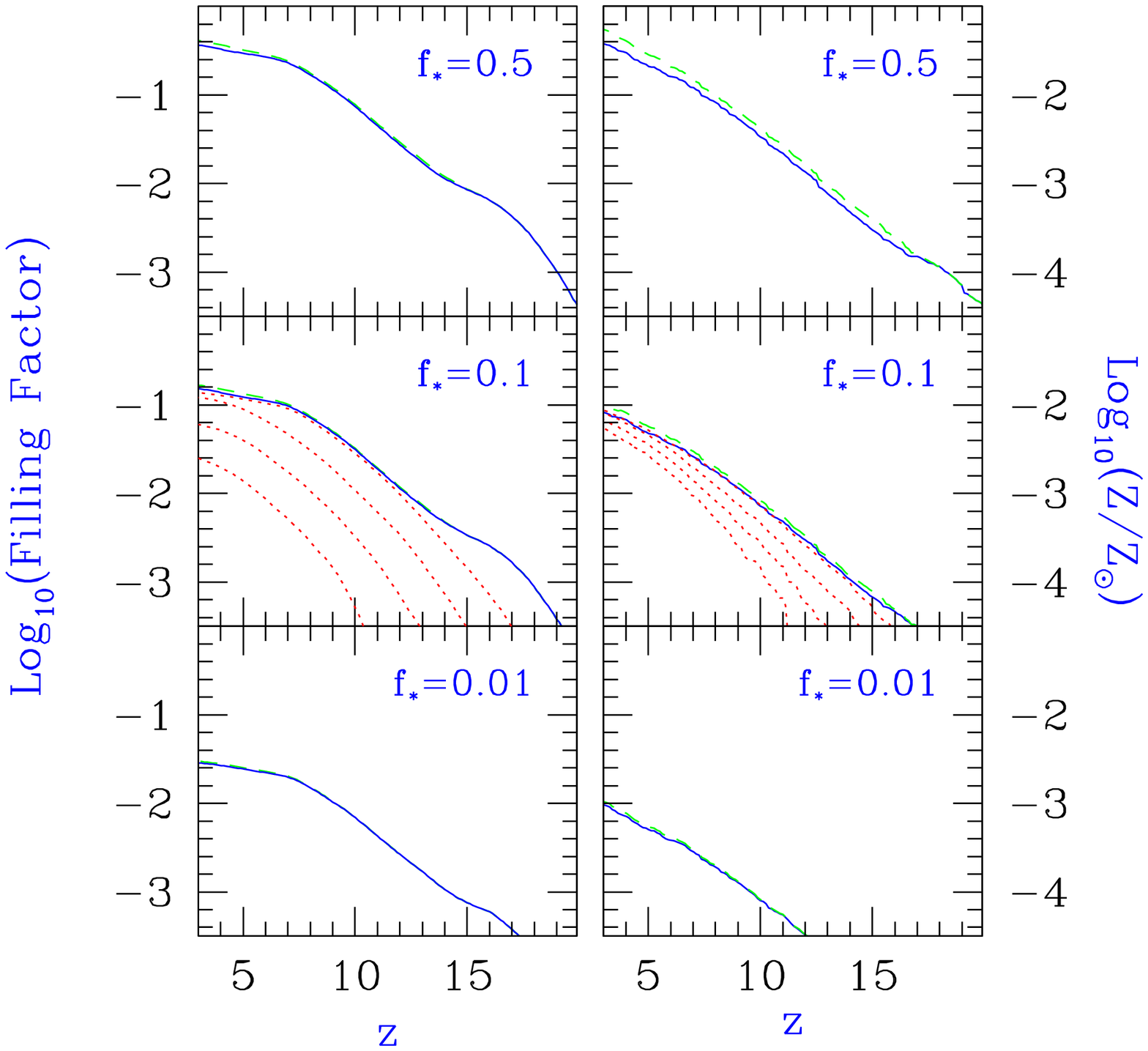,width=4in}}
\caption{\footnotesize {\em Left panels:}
Cosmic metal filling factor from galaxy outflows.  In each panel {\it
solid (dashed)} lines show the filling factor in runs with (without)
the inclusion of baryonic stripping feedback according to eq.\
(\protect\ref{eq:strip}).  Each panel is labeled by the overall star
formation coefficient $f_\star$ and in the $f_\star = 0.1$ case the
{\it dotted} lines represent models with no feedback, but outflows are
only allowed from objects at or above a fixed mass limit.  From top to
bottom these limits are $1.0 \times10^8 M_\odot$, $3.3\times10^8
M_\odot$, $1.1\times10^9 M_\odot$, and $3.6\times10^{9} M_\odot$.
{\em Right panels:} Mass-averaged IGM metallicity as a function of redshift.
Curves are as in the left panels.}
\label{fig:vol}
\vspace{+0.5cm}
\end{figurehere}

The parameters in these models are $f_\star = 0.5, 0.1, 0.01$, $f_w(M)
= 0.3\delta_B(M)/\delta_B(2 \times 10^8 M_\odot)$ with $\delta_B$ as
in eq.\ (\ref{eq:deltaB}), $\nu^{-1} = 136 M_\odot$, $t_{\rm OB}= 33$
Myr, and $f_m = 0.5.$ The first three of these parameters appear in
combination both in the expression for the overall strength of the
winds ($\propto f_\star f_w \nu)$, and the metallicity ($\propto
f_\star \nu$).  On the other hand, $t_{\rm OB}$ has almost no effect
on our results as the relevant times for star formation are small
compared to structure formation times scales.  Thus we can provide a
conservative estimate of the model uncertainties introduced by these
parameters by simply considering a wide range of star formation
efficiencies, and applying a linear shift in the final metallicity to
estimate the effect of varying $f_w$.  Finally, while the mass loading
parameter $f_m$ has little effect on the overall filling factor, it is
important for galaxy feedback, and we consider its impact in detail
\S4.3.

The most obvious, yet perhaps most important feature of Fig.\ 1 is
that the filling factor is always substantially less than unity,
ranging from 3\% to 30\% at $z=3$.  Note that these values are
consistent with the $20\%$ enrichment at $z=4$ found in numerical
simulations by Thacker, Scannapieco, \& Davis (2002), using a model
similar to our $f_\star = 0.1$ case.  The fact that IGM enrichment is
inhomogeneous even in the maximal case in which 50\% of all baryons in
collapsed objects are taken to form stars, however, leads us to an
important conclusion: starburst driven outflows, while an effective
source of metals in overdense regions (SB), are not able to enrich the
IGM in its entirety.  This is true even in the $\Lambda$CDM model
considered in our simulations, in which dwarf galaxies are formed at
very high redshifts, and the baryonic/dark matter ratio is relatively
high, resulting in a large number of stars.

The details of our results depend sensitively on the minimum mass
scale of the galaxies in our simulation, however, which is set by our
minimum virial temperature of $10^4$~K.  In the central panel of this
figure, we plot a series of models in which no feedback as per
eq.\ (\ref{eq:strip}) is imposed, but instead we allow outflows
only from objects above a fixed mass scale.  Both the redshift at
which outflows begin to become important and their overall filling
factor depends closely on this mass.  Thus, while in the run with
$f_\star = 0.1$, outflowing bubbles fill 1\% of the volume at redshift
$\lsim 12$ and reach a final filling factor of $16\%$, excluding all
objects with masses below $1.1 \times 10^9 M_\odot$ shifts these
values to $z \approx 8$ and $6\%$ respectively.  Note that this lower
resolution is similar to that adopted by Aguirre et al.\ (2001a) and
approximately equal to the mass of a single dark matter particle in
the simulations by Cen and Ostriker (1999).

In spite of the sensitivity of metal enrichment to low-mass objects,
its overall dependence on baryonic stripping feedback is weak, as can
be seen by comparing the solid lines in which equation
(\ref{eq:strip}) has been imposed with the dashed lines in which such
feedback from outflows is neglected.  The shape and final value of the
filling factor are extremely similar between such models for all
values of $f_\star$, becoming indistinguishable in many cases. This is
because baryonic stripping can only occur in a perturbation that is
sufficiently nearby and late--collapsing. Then the shock velocity,
$\dot R_s$, is large and the overdense region occupies a large solid
angle, $\omega$, when the outflow reaches it.  Thus the perturbations
succumbing to baryonic stripping correspond to late--forming galaxies
in the most heavily populated regions of space, which have little
effect on the overall filling factor.

The higher bias of suppressed objects can also be seen by comparing
the evolution of the filling factor with the overall mass-averaged IGM
metallicity, plotted in the right panels of Figure \ref{fig:vol}.  In
these panels, the differences between the models with and without
suppression are much more pronounced.  The difference is most apparent
in the $f_\star=0.5$ case, in which the wind velocities are the
highest, and thus the suppression of neighbors is most severe.  In
this case at $z = 3$ the overall metallicities differ by a factor of
1.5 while the difference in volume filling factor is less than a
factor of 1.15.

Note that the mass-averaged metallicity scales almost linearly with
$f_\star$, as this parameter controls the number of stars formed in
each galaxy, and hence the number of supernovae and mass of ejected
metals.  We find that at $z = 3$, $Z \approx 0.1 f_\star$, where this
relation depends on the assumed yield ($2 M_\odot$ per SN, 1/2
ejected), the gas ejected fraction ($50\%$), and the minimum mass
scale in the simulation.  This mass dependence, while sensitive, is
more limited than that of the overall filling factor, as can be seen
by comparing the $f_\star = 0.1$ model with the series of models with
a threshold mass imposed, plotted in the center right panel.  Thus the
model with $T_{\rm vir} \geq 10^4$ K (solid line) and that with a
fixed threshold of $1.1 \times 10^9 M_\odot$ have $z = 3$
metallicities of $0.010~Z_\odot$ and $0.007~Z_\odot$ respectively.

\begin{figurehere}
\centerline{
\hspace{+0.4cm}
\psfig{file=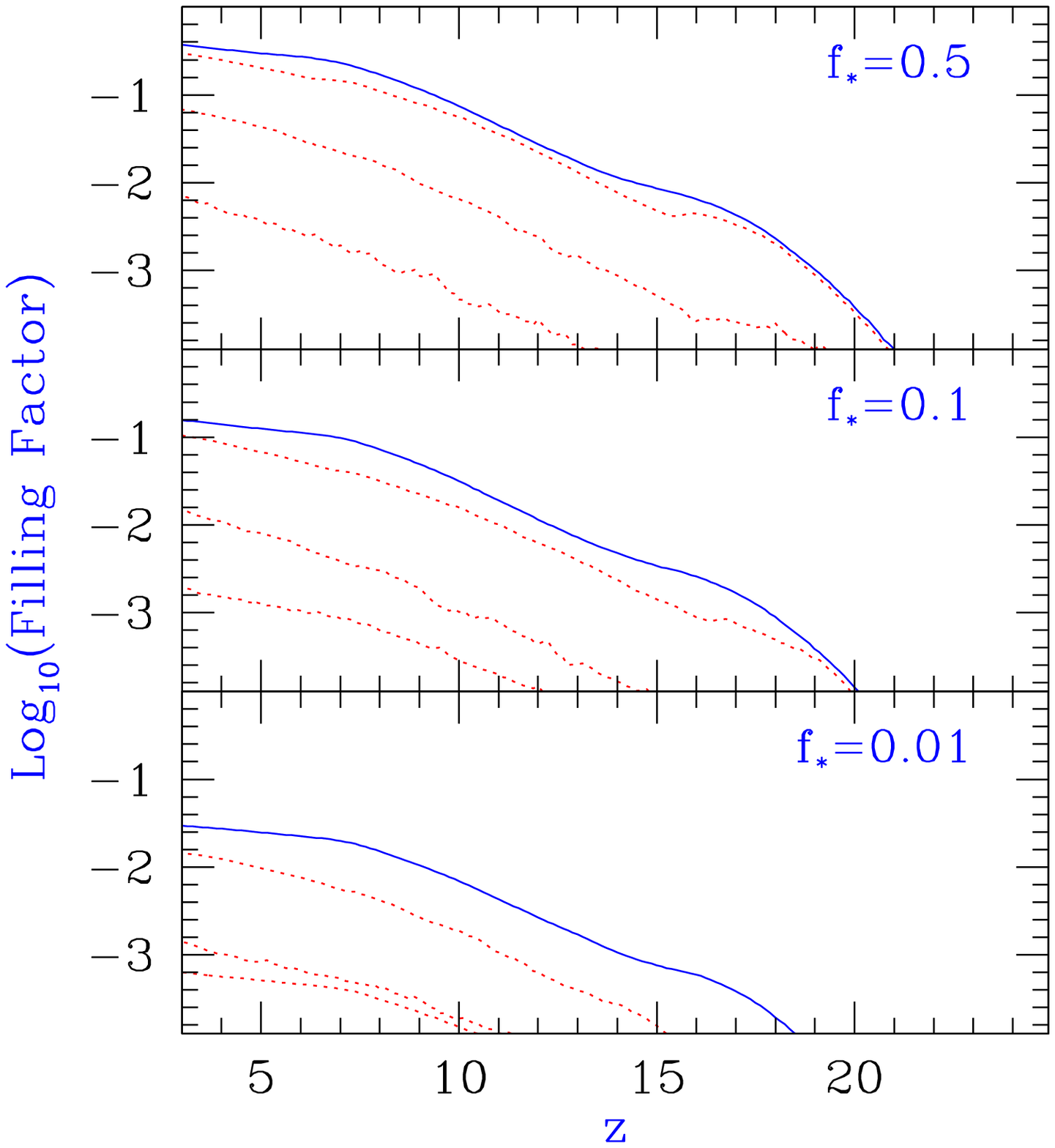,width=4in}}
\caption{\footnotesize Cosmic metal filling factor from galaxy outflows, as a
function of threshold metallicity.  In each panel the solid lines show
the total filling factor of metals, which is indistinguishable from the
filling factor of gas with $Z \ge 10^{-3} Z_\odot$.  
The dotted lines, from top to bottom, show the filling
factor  of gas enriched to values above $(10^{-2}, 10^{-1}, 1) 
Z_\odot$, respectively.}
\label{fig:vzc}
\vspace{+0.2cm}
\end{figurehere}

In Figure \ref{fig:vzc} we examine the distribution of metals in more
detail, by computing the filling factor of gas above a fixed threshold
metallicity.  As the metallicities of individual outflows always
exceed $10^{-3} Z_\odot$ in all models, the filling factor at this
threshold metallicity is indistinguishable from the total filling
factor as shown in Figure \ref{fig:vol}.  

Choosing a higher threshold value, however, leads to more
model-dependent results.  From this point of view, the $f_\star=0.5$
and $f_\star=0.1$ models are quite similar, and in both cases
approximately $50\%$ of the enriched gas has a metallicity greater
than $10^{-2} Z_\odot$, $10\%$ of the enriched gas exceeds $10^{-1}
Z_\odot$, and $1\%$ of the enriched IGM exceeds solar metallicity.
On the other hand, the distribution of metals is much more clumpy in
the weak-star formation ($f_\star = 0.01$) case.  Here, while about
$50 \%$ of the enriched IGM exceeds $10^{-2} Z_\odot$, the volume
fraction of enriched gas with metallicities above $10^{-1} Z_\odot$
and $Z_\odot$ are quite similar, with such high-metallicity gas
occupying approximately $6\%$ and $3\%$ of the enriched volume
respectively.

Our estimates of the IGM metallicity invite comparison with that
recently observed in the low-column density Ly$\alpha$ forest.
Songaila (2001) derived the redshift evolution of $\Omega_\nCIV$ and
$\Omega_\nSiIV$ for a large number of systems (367 and 109 for \CIV
and \SiIV) with \HI column densities $10^{12}\cmm < N_\nHI <
10^{15}\cmm$.  The above quantities can then be translated, modulo the
assumption of the ionization fraction of such species taken to be
equal to 0.5, into a metallicity estimate. The resulting metallicity
evolution is shown by the points in Fig. \ref{fig:song}, with the error
bars corresponding to a 90\% confidence limit. Songaila points out two
striking features of IGM enrichment: (i) the existence of a minimum
metallicity $Z \approx 10^{-3.45} Z_\odot$, already in place at $z=5$,
and (ii) a constancy of metal abundances through the studied redshift
range $z=2 \rightarrow 5$. A direct comparison of our simulations with
these results is possible provided that we filter our metallicity
estimates with the overdensities sampled by Songaila's
experiment. Ricotti, Gnedin \& Shull (2000) showed that a clear
correlation between \HI column and gas overdensity exists, with
$\rho_b/\overline\rho_b\simeq 0.8\,N_{\nHI,13}^{0.7}$; hence, the
$10^{12}\cmm < N_\nHI < 10^{15}\cmm$ range can be translated into the
overdensity range $0.16 \simlt \rho_b/\overline\rho_b \simlt 20$.

\begin{figurehere}
\centerline{
\hspace{+0.4cm}
\psfig{file=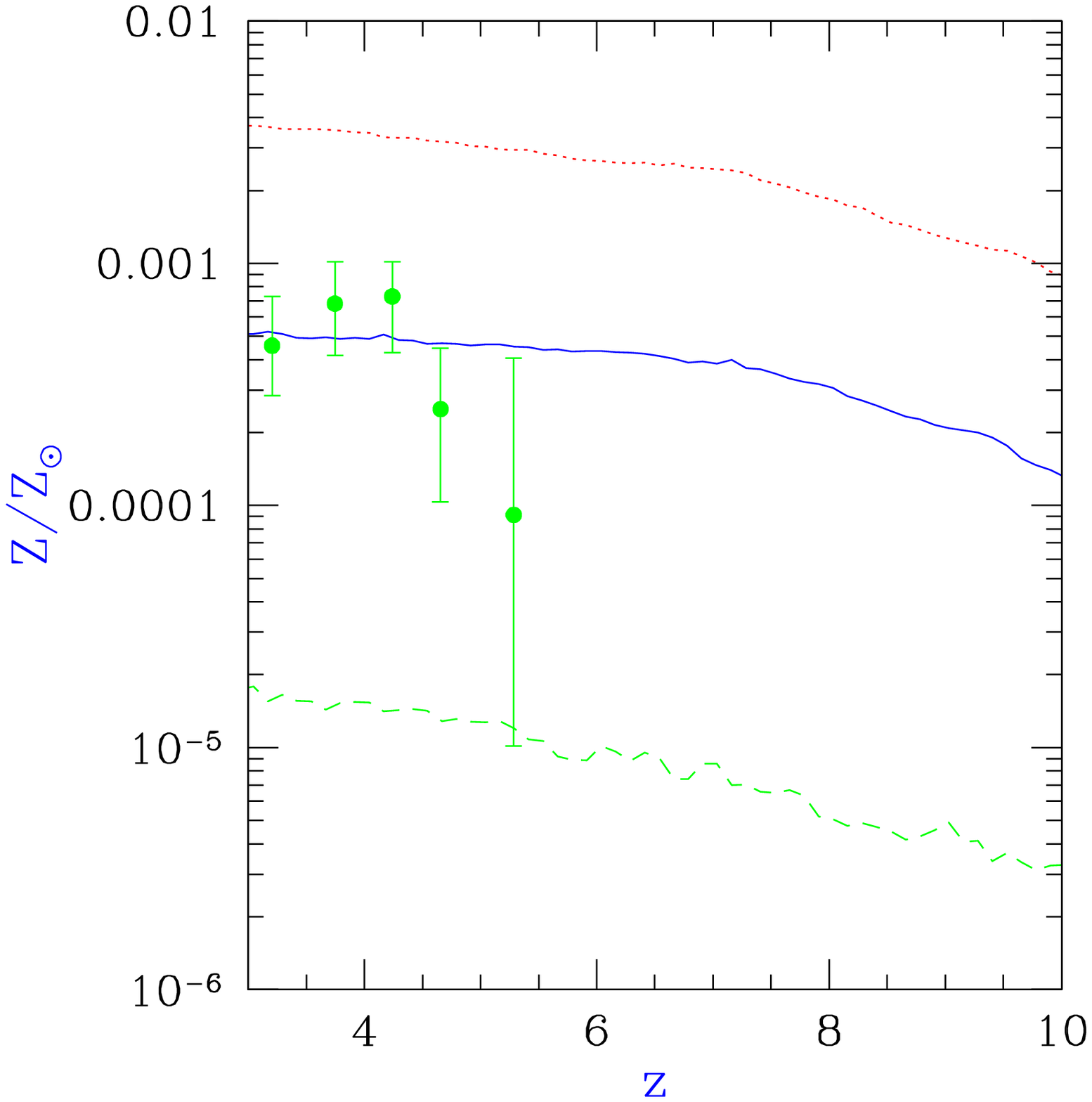,width=4.0in}}
\caption{\footnotesize Metallicity evolution for low density 
($10^{12}\cmm < N_\nHI < 10^{15}\cmm$) IGM. Points are the data of Songaila (2001); dotted, solid,
and dashed lines
are the predicted values for $f_\star =0.5,0.1,0.01$ respectively.} 
\label{fig:song}
\vspace{+0.5cm}
\end{figurehere}

Adopting the smallest filter scale and associating linear and
nonlinear overdensities according to eq.\ ($\ref{eq:sphere}$), we have
searched through our simulations and computed a mass averaged
metallicity of cells that are not surrounded by a collapsed
perturbation and in which $0.8 \simlt \rho_b/\overline\rho_b \simlt
20,$ roughly corresponding QSO absorption systems with $10^{13}\cmm <
N_\nHI < 10^{15}\cmm$.  We adopt this lower bound as our mapping
between linear and nonlinear overdensities becomes unreliable for smaller
values of $\delta$.  These results are superposed on the data points in
Fig.\ \ref{fig:song} for the usual three values of $f_\star$. Although
no fitting parameter has been adjusted to produce these curves, the
resemblance to the data, both in terms of amplitude and flat shape, is
noticeable. It is only at redshifts above $z=8$ that the curve starts
to bend downwards.  The physical interpretation for the existence of
this metallicity floor is simple: metal enrichment is dominated by
low-mass ($10^8 - 10^9 M_\odot$) objects for which metal
ejection/transport occurs with highest efficiency (see discussion in
Ferrara, Pettini \& Shchekinov and eq.\ \ref{eq:deltaB}). As the
nonlinear mass scale increases, larger but intrinsically less
efficient galaxies become the most numerous population. However, they
contribute only marginally to metal enrichment of the Ly$\alpha$
forest as synthesized metals are predominantly trapped by their
potential wells. This explains both the shape and amplitude of the
data, provided $f_\star \approx 0.1$, a value that is broadly
consistent with other arguments based on cosmic star formation rates
(Ciardi \etal 2000; Thacker, Scannapieco, \& Davis 2002; Barkana
2002).

\subsection{Population III Objects}

In order to study the impact of Pop III objects on our results, we
have conducted a number of comparison simulations in which we allow
objects with virial temperatures $ < 10^4$~K to form, but with a
lower overall star formation efficiency.  In these cases
we took $f_{\star PopIII} = 0.1 f_\star$ in objects with $T < 10^4$~K. This
is physically motivated by the lower efficiency of molecular hydrogen
cooling, which reduces the number of cooled baryons available to form
stars.  For each value of $f_\star$ we conducted three such comparison
runs: one with the same overall resolution as the fiducial case (a 4
$h^{-1}$ Mpc comoving box with objects with nine mass scales ranging
from $3.0 \times 10^7\,\msun$ to $4.3 \times 10^{11}\,\msun$), one in
a 2.75 $h^{-1}$ Mpc comoving box with the nine mass scales now ranging
from $1.0 \times 10^8 \msun$ to $1.3 \times 10^{11} \msun$, and one in
a 1.75 $h^{-1}$ Mpc comoving box with masses ranging from $3.0 \times
10^6 \msun$ to $4.3 \times 10^{11} \msun$.  Note that in the last of
these simulations the smallest objects have virial temperatures $\leq
5,000$ K at redshifts below 20, and are thus not only well below the
$10^4$ K atomic cooling limit, but also below the $5,000$ K limit for
``efficient'' molecular cooling at which the cooling time exceeds the
free-fall time (MFR).

The filling factor and overall metallicities in the comparison runs
are shown in Figure \ref{fig:conv}.  Here we restrict our analysis to
redshifts above 5, so that our linear approach remains accurate at the
smallest mass scales.  In this figure we see that even in the case
with the highest resolution, Pop III objects are only able to increase
the metal filling factor and metallicity of the IGM to values that
are a factor of 1.4 times greater than those found in the standard
runs.  Similarly, these objects have only a small impact on the
redshift at which outflows fill $1\%$ of the volume, pushing this
value from 12 to 14 in the $f_\star = 0.1$ case.  From this
comparison, it is clear that while our results are sensitive to
imposing a mass limit above $T_\vir = 10^4$ K, they are relatively
insensitive to the inclusion of Pop III objects, which rely on ${\rm H}_2$
line cooling for collapse.

\begin{figurehere}
\centerline{
\hspace{+0.4cm}
\psfig{file=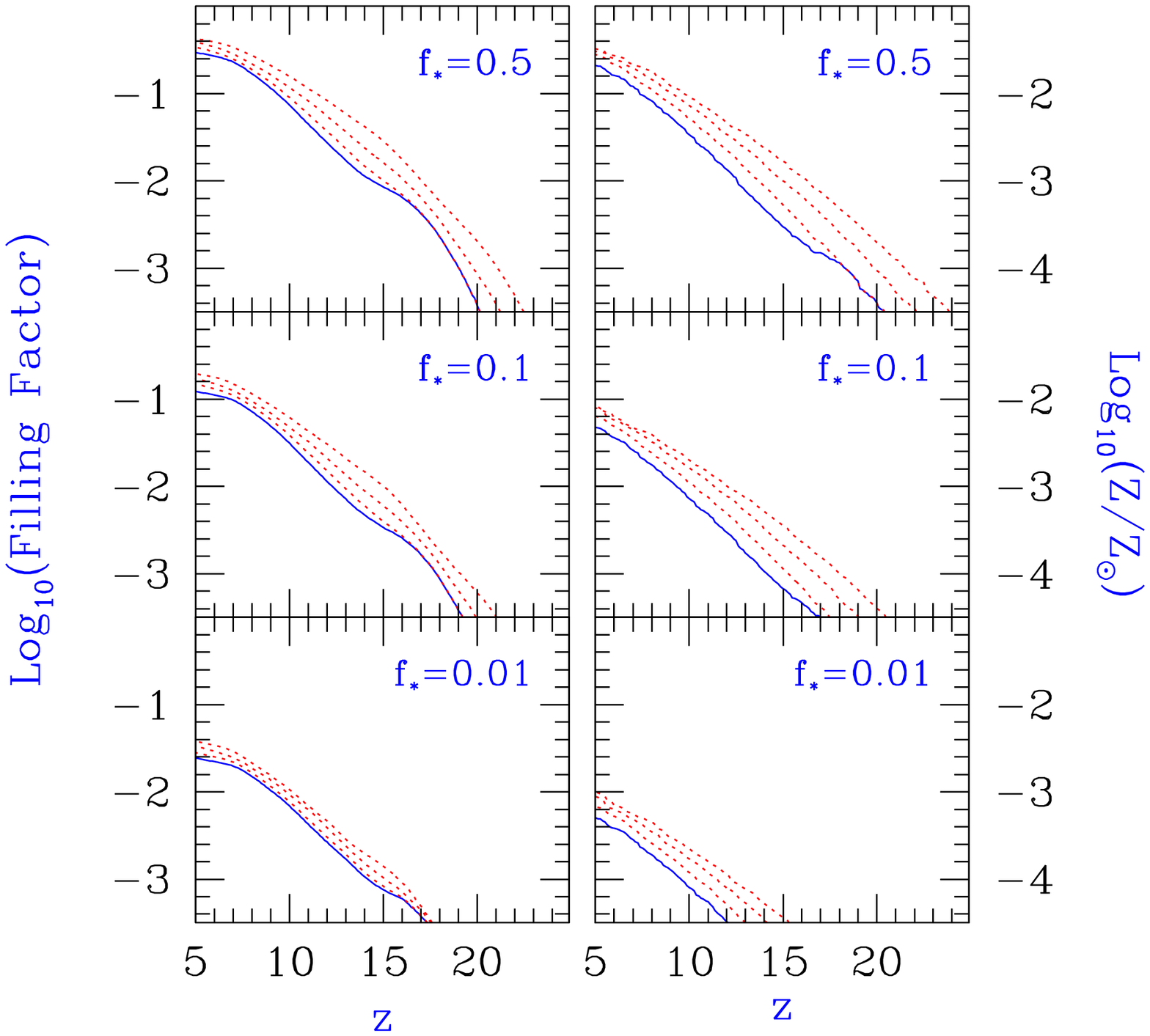,width=4in}}
\caption{\footnotesize Impact of Pop III objects.  In each panel the
{\it solid} lines show the filling factor and total metallicity in runs
with a minimum mass scale of $3.0 \times 10^7$ $M_\odot$ and a minimum
$T_{\rm vir} = 10^4$ K as in Fig. 1.  
The {\it dotted} lines are taken from runs which
include star formation in objects $T_{\rm vir}
< 10^4$ K as described in the text.  From top to bottom the minimum
mass scale in these simulations is $3.0 \times 10^6$ $M_\odot$, $1.0
\times 10^7$ $M_\odot$, and $3.0 \times 10^7$ $M_\odot$,
corresponding to virial temperatures of 5,000, 12,000, and 26,000~K 
at $z=20$ and 2,800, 6,000, and 13,000~K at $z=10$.}
\label{fig:conv}
\vspace{+0.5cm}
\end{figurehere}

\subsection{Feedback}

In Figure \ref{fig:ff} we explore in greater detail the
relationship between the cosmic filling factor and baryonic stripping.
While the suppression of galaxy formation has little impact on the
cosmic metal filling factor, these quantities are clearly related, as
both are dependent on the number density and strength of galaxy
outflows.  In this plot we quantify baryonic stripping by counting all
galaxies with masses above $3.2 \times 10^8 M_\odot$, excluding
the objects at the smallest two mass scales, as at low redshifts their
virial temperatures fall below the $10^4$ K limit.  In this case, the
majority of suppressed objects are slightly larger dwarf
galaxies with masses of a few times $10^9 M_\odot$ (SB; Scannapieco,
Thacker, \& Davis 2001).

In the upper panel of this figure, we see that while the details of
outflow generation and propagation introduce a large scatter,
widespread IGM enrichment is accompanied by a significant level of
baryonic stripping in all cases.  Thus even given the wide range of
$f_\star$ values considered, all models and redshifts in which $2\%$
of the IGM is enriched show a suppression of galaxies $\gsim 20 \%$.
In the $f_\star = 0.1$ case, which is most consistent with QSO
observations, approximately half of the objects with masses at or
above $3.2 \times 10^8 M_\odot$ are suppressed by this mechanism by $z
= 3$.

These results are sensitive however, to the mass-loading parameter
$f_m$, as the suppression of galaxy formation is primarily due to the
momentum carried by the winds (see eg.\ Scannapieco, Ferrara, \&
Broadhurst 2000; Scannapieco, Thacker, \& Davis 2001), which scales as
$E_b^{1/2} M_s^{1/2}.$ Thus increasing the fraction of gas swept into
the shells increases galaxy suppression in models with the same
kinetic energy input.  In the lower panel of Figure \ref{fig:ff}, we
compare our fiducial $f_\star=0.1$, $f_m=0.5$ model with enhanced
(depressed) feedback models in which $f_m = 1.0$ ($f_m = 0.25$).
Although changes in $f_\star$ result in a simultaneous increase of both
the filling factor and suppression factor of dwarf galaxies, changing
$f_m$ shifts the suppression factor while having little effect on the
overall filling factor.  Thus the filling factors in these models are
all between $14\%$ to $18\%$, while the fraction of suppressed dwarf
galaxies varies from $35\%$ to $60\%$.

\begin{figurehere}
\centerline{
\psfig{file=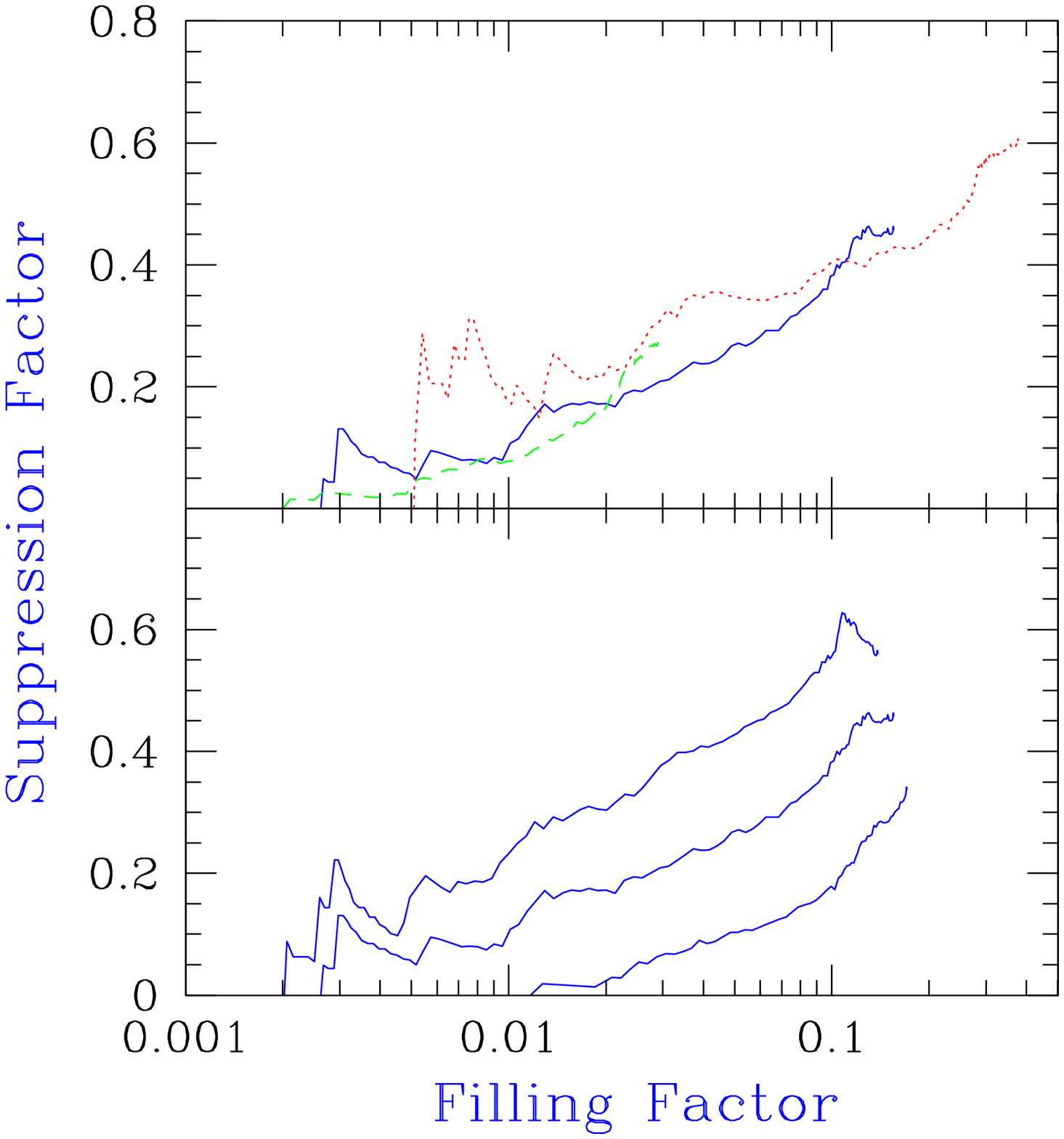,width=4in}}
\caption{\footnotesize {\em Top: }Fraction of galaxies suppressed by baryonic
stripping as a function of the overall metal filling factor.
The {\it dotted, solid,} and {\it dashed} curves correspond to
models in which $f_\star$ is set to $0.5$, $0.1$, and $0.01$ 
respectively. {\em Bottom:} Impact of $f_m$ on the suppression of galaxy
formation.  From top to bottom $f_m = 1.0, 0.5,$ and $0.25$, with 
$f_\star= 0.1$ in all cases.}
\label{fig:ff}
\vspace{+0.5cm}
\end{figurehere}

As a general remark about this type of feedback, we note that the gas
is swept from the potential well is unlikely to be reaccreted by this
object.  Instead, this gas be available to other collapsing objects,
which therefore may be characterized by higher baryon-to-dark matter
ratios.  Clearly, in order to draw more quantitative conclusions a
dedicated hydrodynamical study of the details of baryonic stripping is
required.

\section{Conclusions}

In this work we have studied the metal enrichment of the IGM by
outflows in a $\Lambda$CDM model of structure formation and its
feedback on the formation of galaxies.  Adopting a linear peaks model
of the spatial distribution of forming objects, and a detailed
one-dimensional model of wind propagation, we have determined the
overall filling factor as a function of redshift and its relationship
with the baryonic stripping of protogalaxies.

While the star formation efficiency of high-redshift galaxies is
largely unknown, we are nevertheless able to place useful constraints
on the filling factor, enrichment redshift, and overall mass-averaged
metallicity in such models.  Choosing a range of star formation
efficiencies between $f_\star = 0.50$ and $f_\star = 0.01$, we find
that at least $3\%$ and at most $30\%$ of the IGM is enriched to a
level exceeding $10^{-3} Z_\odot$ by redshift $z=3$.  In all cases,
the majority of this enrichment occurs relatively early, $5 \lsim z
\lsim 12 $, and leads to mass-averaged cosmic metallicities that range
from $0.001 Z_\odot$ to $0.05 Z_\odot$, for star formation
efficiencies $0.01 < f_\star < 0.5$, respectively.  The mass-averaged
metallicity scales roughly linearly with this quantity: $Z \approx 0.1
f_\star Z_\odot.$

Our model can satisfactorily reproduce the constant ($Z\approx 3.5
\times 10^{-4} Z_\odot$) metal enrichment of the low column density
Ly$\alpha$ forest up to $z=5$ derived by Songaila (2001), which is
likely to be caused by the decreasing efficiency of metal loss from
larger galaxies. This comparison strongly favors star formation
efficiencies in a narrow range around 10\%, essentially excluding the
$f_\star = 0.5$ and $f_\star = 0.01$ models.  As the formation of
stars in Pop III objects is relatively inefficient, the inclusion of
these objects has only a secondary effect on our results: increasing
the mass-averaged metallicity and filling factors by at most a factor
of 1.4, and moving the dawn of the enrichment epoch to $z \approx 14$
at the earliest.

While all the models studied display suppression of galaxy formation
due to outflows ram-pressure stripping the gas out of
pre-virialized protogalaxies, this mechanism has only a minor impact
on the overall filling factor as it occurs only in the densest and
most polluted regions of space.  Nevertheless, after fixing $f_m$, a
general relationship between the filling factor and the suppression
factor of galaxies exists, at all $f_\star$ values.  All models and
redshifts at which $2\%$ of the IGM is enriched show a greater than
$20\%$ suppression of galaxies.  In the case that is most consistent
with QSO observations, in fact, half the galaxies are suppressed due
to baryonic stripping.  Thus the relative quiescence of the Ly$\alpha$
forest at lower redshifts is likely to belie a violent epoch of early
outflows and enrichment.

\acknowledgments 

Support for this project was provided by NASA through ATP grant NAG5--4236   
and LTSA grant NAG5--11513, and by a B. Rossi visiting fellowship at the 
Observatory of Arcetri (PM).  ES was supported in part by an NSF MPS-DRF
fellowship.  AF and PM also acknowledge the support of the EC RTN
network ``The Physics of the Intergalactic Medium.''

\end{document}